\documentclass{article}

\usepackage{PRIMEarxiv}
\usepackage{cite}
\usepackage{amsmath,amssymb,amsfonts}
\usepackage[bb=boondox]{mathalfa}
\usepackage{algorithmic}
\usepackage{graphicx}
\usepackage{textcomp}
\usepackage{xcolor}
\usepackage{lipsum}
\usepackage{mathtools}
\usepackage{soul}
\usepackage{comment}
\usepackage[utf8]{inputenc} 
\usepackage[T1]{fontenc}    
\usepackage{hyperref}       
\usepackage{url}            
\usepackage{booktabs}       
\usepackage{amsfonts}       
\usepackage{nicefrac}       
\usepackage{microtype}      
\usepackage{lipsum}
\usepackage{fancyhdr}       
\usepackage{graphicx}       
\graphicspath{{media/}}     
\DeclarePairedDelimiter\floor{\lfloor}{\rfloor}
\title{Fast classical simulation of qubit-qudit hybrid systems
}

\author{
Haemanth Velmurugan\\
University of California\\
San Diego, CA\\
USA\\
\And
Arnav Das\\
India Internet Foundation\\
Kolkata\\
India\\
\And
Turbasu Chatterjee\\
Virginia Tech\\
Blacksburg, VA\\
USA\\
\And
Amit Saha\\
Institut National de Recherche en Informatique et en Automatique (Inria)\\
\& École Normale Supérieure (ENS), PSL University\\
Paris\\
France\\
\And
Anupam Chattopadhyay\\
School of Computer Science and Engineering\\ 
Nanyang Technological University\\
Singapore\\
\And
Amlan Chakrabarti\\
A. K. Choudhury School of Information Technology\\
University of Calcutta\\
Kolkata\\
India\\
}

\begin{document}
\maketitle

\begin{abstract}
Simulating quantum circuits is a computationally intensive task that relies heavily on tensor products and matrix multiplications, which can be inefficient. Recent advancements, eliminate the need for tensor products and matrix multiplications, offering significant improvements in efficiency and parallelization. Extending these optimizations, we adopt a block-simulation methodology applicable to qubit-qudit hybrid systems. This method interprets the statevector as a collection of blocks and applies gates without computing the entire circuit unitary. Our method, a spiritual successor of the simulator QuDiet \cite{Chatterjee_2023}, utilizes this block-simulation method, thereby gaining major improvements over the simulation methods used by its predecessor. We exhibit that the proposed method is approximately 10$\times$ to 1000$\times$ faster than the state-of-the-art simulator for simulating multi-level quantum systems with various benchmark circuits. 
\end{abstract}


\section{Introduction}
Quantum computing has rapidly advanced, becoming a significant focus of contemporary research. The application of quantum algorithms is increasingly prevalent across various scientific and technological domains. The increasing complexity and size of quantum algorithms \cite{dalzell2023quantumalgorithmssurveyapplications}, are essential for using them in more application-oriented use cases and require the development of larger quantum circuits beyond the 2-dimensional qubit realm \cite{sahapra}. The growing need for higher-dimensional systems necessitates advanced classical simulators for multidimensional quantum circuits \cite{Wang_2020}. These play a crucial role in developing and testing quantum algorithms, providing a vital platform for researchers to explore and refine quantum computing techniques before deployment on actual scalable quantum hardware \cite{Ringbauer_2022}.


\paragraph{Motivation}
Despite the existence of many state-of-the-art simulators like Qiskit \cite{qiskit2024} and Cirq \cite{cirq_developers_2024_11398048}, Jet \cite{Vincent_2022}, MQT Qudits \cite{mato2024mqtquditssoftwareframework}, QuDiet provides a user-friendly interface for simulating hybrid qubit-qudit operations.
While QuDiet is a first-of-its-kind, user-friendly classical simulator for hybrid quantum systems, there is significant scope for improvements. QuDiet also relies on tensor products and matrix multiplications to simulate quantum circuits. In contrast, \textit{q}H\textit{i}PSTER \cite{smelyanskiy2016qhipster} introduces and QuEST\cite{jones2019quest} further developed, a direct evolution approach that entirely eliminates both tensor products and matrix multiplications for binary state-space simulations, offering opportunity for parallelization. Studies from Microsoft\cite{jaques2022leveraging}, have further optimized this approach by using hash maps for efficient sparse representation of the statevectors producing some promising results in comparison to other existing state-of-the-art methods for qubit-only systems, as exhibited in Table \ref{tab:qis_cir_qus_comp}. 

\begin{table}[!h]
\centering
\caption{The execution time (in seconds) of random circuits with Clifford+T gates using different classical simulators (CNOT-depth of all circuits was fixed to be 500 and the width of a quantum circuit is the number of qubits).}
\label{tab:qis_cir_qus_comp}
\begin{tabular}{|l|l|l|l|}
\hline
        Width  & Qiskit & Cirq & Quest \\ \hline
        2  & 0.1076 & 0.2017 & 0.000001 \\ \hline
        3  & 0.1262 & 0.1213 & 0.040909 \\ \hline
        4  & 0.1551 & 0.1367 & 0.000001 \\ \hline
        5  & 0.1893 & 0.1686 & 0.000001 \\ \hline
        6  & 0.2104 & 0.2032 & 0.000001 \\ \hline
        7  & 0.2427 & 0.2435 & 0.000101 \\ \hline
        8  & 0.2712 & 0.2779 & 0.001818 \\ \hline
        9  & 0.3005 & 0.3208 & 0.006869 \\ \hline
        10  & 0.3247 & 0.3681 & 0.010202 \\ \hline
        11  & 0.3707 & 0.4097 & 0.014141 \\ \hline
        12  & 0.4968 & 0.4816 & 0.025657 \\ \hline
        13  & 0.557 & 0.7227 & 0.052525 \\ \hline
        14  & 0.768 & 0.8397 & 0.10404 \\ \hline
        15  & 0.877 & 1.1243 & 0.21899 \\ \hline
        16  & 1.048 & 1.677 & 0.461515 \\ \hline
        17  & 2.092 & 2.7225 & 0.990606 \\ \hline
        18  & 2.553 & 5.0096 & 2.049293 \\ \hline
        19  & 3.514 & 11.736 & 4.297551 \\ \hline
        
\end{tabular}
\end{table}


Although it is known that these optimizations may bring improvements to qubit-qudit simulations, they also pose a significant challenge in the implementations of such optimizations in the qudit realm. This is mainly due to the very large dimensions of the qudit vectors and the operators. In this article, we adopt a block simulation methodology, which extends the concept of direct evolution as discussed in QuEST\cite{jones2019quest} by Jones et.al., to simulate higher and mixed-dimensional quantum systems faster.

\section{Proposed Methodology}
Here, we propose the block simulation method for any finite-dimensional quantum system. Initially, the method proposed is based on the idea that any arbitrary single-qudit gate can be applied to a quantum state by manipulating the statevector directly, later on, we extend this idea to multi-qudit gates. The action of every single-qudit gate is localized to certain blocks of amplitudes and one can apply a given unitary by identifying the blocks and modifying them in place. We consider three cases of single qudit gates – phase gates (diagonal unitaries), permutation gates, and others (which result in superposition). \\

Consider a multi-dimensional qudit system of $n$ qudits denoted as $q_0, q_1, \hdots, q_{n-1}$ with dimensions $d_0, d_1, \hdots, d_{n-1}$. We consider $q_0$ as the least significant qudit and $q_{n-1}$ as the most significant qudit respectively. The statevector of the system can be expressed as \\
\begin{center}
    $|\psi\rangle = \begin{pmatrix} \alpha_0 \\ \alpha_1 \\ \vdots \\ \alpha_{D-1} \end{pmatrix}$ where $D = \prod_{i=0}^{n-1}d_i.$
\end{center}

Each element of the statevector $\alpha_j$ corresponds to the amplitude of the basis state $|j\rangle$ respectively. The binary representations of the basis states follow a pattern that can be exploited to identify blocks of states. Given a single-qudit unitary $U$ acting on qudit $q_i$, the statevector can be viewed as a collection of blocks such that the $i^{th}$ qudit $q_i$ takes a fixed value within any given block as shown below.

\begin{center}
$\hspace{2.2cm} q_{n-1}q_{n-2}\hdots q_{i+1}\textcolor{red}{q_{i}}q_{i-1}\hdots q_{1}q_0$ \\
$|\psi\rangle = \begin{pmatrix} \boxed{} \\  \boxed{}  \\ \vdots \\  \boxed{} \\  \boxed{} \\  \boxed{} \\ \vdots \\ \\ \vdots \\ \boxed{} \\  \boxed{} \\ \vdots \\  \boxed{} \\ \vdots \end{pmatrix} \begin{matrix} \boxed{ \begin{matrix} 00\hdots0\textcolor{red}{0}0\hdots00 \\ 00\hdots0\textcolor{red}{0}0\hdots01 \\ \vdots \\ 00\hdots0\textcolor{red}{0}(d_{i-1})\hdots(d_1-1)(d_0-1) \end{matrix}} \\  \boxed{ \begin{matrix} 00\hdots0\textcolor{red}{1}0\hdots00 \\ 00\hdots0\textcolor{red}{1}0\hdots01 \\ \vdots  \end{matrix}}  \\ \vdots \\  \boxed{ \begin{matrix} 00\hdots0\textcolor{red}{(d_i-1)}0\hdots00 \\ 00\hdots0\textcolor{red}{(d_i-1)}0\hdots01 \\ \vdots  \end{matrix}} \\ 00\hdots1\textcolor{red}{0}0\hdots00 \\ \vdots \end{matrix}$ 
\end{center}

Viewing the statevector as blocks would simplify the application of the gates for the different gate categories as described.


\subsection{Phase Gates}

A diagonal unitary $U$ of a phase gate, when applied to qubit $q_i$ would only change the amplitudes of the basis states already present in the statevector. These unitaries preserve the states in the statevector and do not introduce new states or superpositions. The only task here is to scale the non-zero amplitudes in the statevector.

The action of U on qubit $q_i$ in state $|j\rangle$ is equivalent to multiplying its amplitude $\alpha_j$ by the corresponding diagonal entry $U(j,j)$. As seen in the diagram, all states in the first block of the statevector have qudit $q_i$ in state $|0\rangle$. The action of $U$ on this block can be achieved by multiplying the entire block by the scalar $U(0,0)$. Similarly, the second block, which has the qudit $q_i$ in state $|1\rangle$ is multiplied by the scalar $U(1,1)$. This process is repeated for all blocks to arrive at the new statevector. Examples of gates in this category include the generalized $Z$ gate and its derivatives – the $T$ gate and $S$ gate.

\subsection{Permutation Gates}

This type of gate maps every basis state to another unique basis state. The most common example is the generalized NOT gate $X_{+a}^d$ which performs the following mapping of single-qudit states $|j\rangle \rightarrow |(j+a)$ mod $d\rangle$.

Since block $k$ of the statevector contains basis states with qudit $q_i$ in state $|k\rangle$, the $X_{+a}^{d_i}$ gate acting on qudit $q_i$ essentially maps block $k$ to block $(k+a)$ mod $d_i$. This block transformation is localized to every consecutive $d_i$ block whose corresponding members differ only in the value of qudit $q_i$ and can be implemented using a cyclic rotation of the $d_i$ blocks by $a$ steps. We repeat this $a$-block rotation of every $d_i$ block to yield the modified state.

\subsection{Superposition gates}

The last category of gates is the one that introduces new states through superpositions. A well-known example is the generalized Hadamard gate. To understand the working of these gates, we introduce the notion of equivalent classes of basis states under the operation of these gates as shown in Table~\ref{tab:eq_classes}. Each class consists of $d_i$ elements which differ only in the value of qudit $q_i$ to which the unitary is applied.

\begin{table}[!h]
    \caption{Equivalent classes of basis states under superposition unitaries.}
    \centering
    \begin{tabular*}{\linewidth}{|p{0.2\linewidth}|p{0.2\linewidth}|c|p{0.472\linewidth}|}
         \hline Class 1 & Class 2 & $\hdots$ & Class N  \\
         \hline
         $00\hdots\begingroup\color{red}0\endgroup\hdots00$ & $00\hdots\begingroup\color{red}0\endgroup\hdots01$ & $\hdots$ & $(d_{n-1}-1)(d_{n-2}-1)\hdots\begingroup\color{red}0\endgroup\hdots(d_1-1)(d_0-1)$ \\
         \hline
         $00\hdots\begingroup\color{red}1\endgroup\hdots00$ & $00\hdots\begingroup\color{red}1\endgroup\hdots01$ & $\hdots$ & $(d_{n-1}-1)(d_{n-2}-1)\hdots\begingroup\color{red}1\endgroup\hdots(d_1-1)(d_0-1)$ \\ 
         \hline
         $\vdots$ & $\vdots$ & $\ddots$ & $\vdots$ \\ 
         \hline
         $00\hdots\begingroup\color{red}(d_i-1)\endgroup\hdots00$ & $00\hdots\begingroup\color{red}(d_i-1)\endgroup\hdots01$ & $\hdots$ & $(d_{n-1}-1)(d_{n-2}-1)\hdots\begingroup\color{red}(d_i-1)\endgroup\hdots(d_1-1)(d_0-1)$ \\
         \hline
    \end{tabular*}
    \label{tab:eq_classes} 
\end{table} 

Consider the generalized Hadamard gate acting on qudit $q_i$. It can be easily verified that the action of the gate on a basis state within a class will only result in a superposition of all states in that class. This is applicable to all basis states within each class and the different classes are independent to each other under the generalized Hadamard operation. The idea holds for any other generic superposition unitary as well.

Class $1$, as shown in Table~\ref{tab:eq_classes}, consists of all the first elements in the first $d_i$ blocks of the statevector. Class $2$ corresponds to all the second elements in $d_i$ blocks. In general, every group of block size = $\prod_{t=0}^{i-1}d_t$  classes contain the corresponding elements from the corresponding $d_i$ blocks. We can therefore find the classes of basis states from the statevector amplitudes. 

In order to apply the unitary $U$ on the input state, we first identify all basis states with non-zero amplitudes in a given class. Each of these basis states will correspond to a particular column of the unitary $U$ that defines its transformation under $U$. In general, the basis state with qudit $q_i$ in state $|j\rangle$ is transformed as dictated by the $j^{th}$ column of $U$. Thus, we can select columns corresponding to non-zero basis states in a given class, perform the scalar product of the columns with the corresponding state amplitudes, and find the sum of these column vectors. Each element in the vectors represents the amplitude of the corresponding basis state in the given class and the vector addition produces the final amplitude of each basis state under the unitary transformation $U$.  The final amplitudes are then stored back in the statevector at the respective locations. 

The proposed method depends on efficiently identifying the blocks and the number of repetitions of these blocks. Based on the basic idea of enumeration, it can be seen that for a qudit $q_i$, the block size is the product of the dimensions of all qudits less significant than the target qudit in the system, which is $\prod_{t=0}^{i-1}d_t$ and the number of repetitions of these $d_i$ block groups is the product of the dimensions of the more significant qudits, which is $\prod_{t=i+1}^{n-1}d_t$. 

\subsection{Controlled and Multi-Controlled Gates}

We can implement multi-controlled gates in a similar fashion by creating a new statevector that contains all the amplitudes corresponding to basis states satisfying the control conditions. The single-qudit unitary is then applied directly on the target using this new statevector. The block size of the target in the new statevector can be found by factoring out the dimensions of all controls that are less significant than the target and the number of repetitions of the blocks is obtained by factoring out the more significant control qudits compared to the target. There are two main advantages here. Firstly, multi-controlled gates can be implemented directly without decomposing them to $1$-qudit and $2$-qudit gate primitives. Secondly, multi-controlled gates whose control conditions are not satisfied are not executed since we select only those basis states that satisfy the control conditions before applying the gate. However, identifying the state amplitudes satisfying all the control conditions requires checking every single index in the statevector and we do not yet know a way to optimize this condition checking.

We consider the application of a multi-controlled unitary $MCU$ on the target qudit $q_i$ from a system of $n$ qudits of dimensions $d_0,d_1,…,d_{n-1}$. We would need two lists specifying the control qudits and the control values. In order to apply the unitary $U$ on target $q_i$, our approach is to construct a partial statevector by extracting only those basis states of the system that satisfy the control conditions. Once we have the partial statevector, we apply the unitary $U$ on target $q_i'$ in the new system corresponding to the target $q_i$ in the original system, obtained by tracing out all the control qudits. 

The brute force approach of iterating through every basis state, and checking if it satisfies all $M$ control conditions takes time of the order of $DM$ where $D=\prod_{i=0}^{n-1}d_i$  and is especially slow for sparse implementations. 

We formally describe the underlying mathematical problem behind our approach. Let the block size of control qudit $q_k$ be denoted by $b_k$ and its control value is $v_k$. A given basis state $|i\rangle$ satisfies the control condition for qudit $q_k$ if 
$\floor*{\frac{i}{b_k}}\mod{d_k}=v_k$, which checks that the qudit $q_k$ is in the state $|v_k\rangle$ in the $n$-qudit basis state $|i\rangle$. This condition can equivalently be expressed as $\floor*{\frac{i}{b_k}}=v_k\mod{d_k}$. Our problem could then be reduced to the following system of equations $ \floor*{\frac{i}{b_k}}=v_k\mod{d_k}  $

This system of equations seems very similar to the Chinese Remainder Theorem’s problem statement \cite{gauss1966disquisitiones}, however, our system has the following differences:
\begin{itemize}
    \item It is not guaranteed that all the divisors $d_k$  are pairwise co-prime – This could be addressed by using the Extended Chinese Remainder Theorem.
    \item The LHS of the expression is no longer a constant value $i$ but $\floor*{\frac{i}{b_k}}$ which varies for every equation in the system.
    \item There are always more than $1$ possible solutions for this system (unless all qudits in the system act as a control implying that there could be no target qudit, which is an impossible case).
\end{itemize}
 
We do not yet know of a way to solve the above system efficiently for more than $1$ control qudits, given the above characteristics. Obtaining a solution to this system could greatly improve the runtime of our simulator for multi-controlled gates, which remains a future scope of this work. For better understanding of our proposed approach, we have exhibited an example circuit with our proposed simulation method. 

\subsection{Example Circuit}
\begin{figure}[!h]
    \centering
    \includegraphics[width=.5\linewidth]{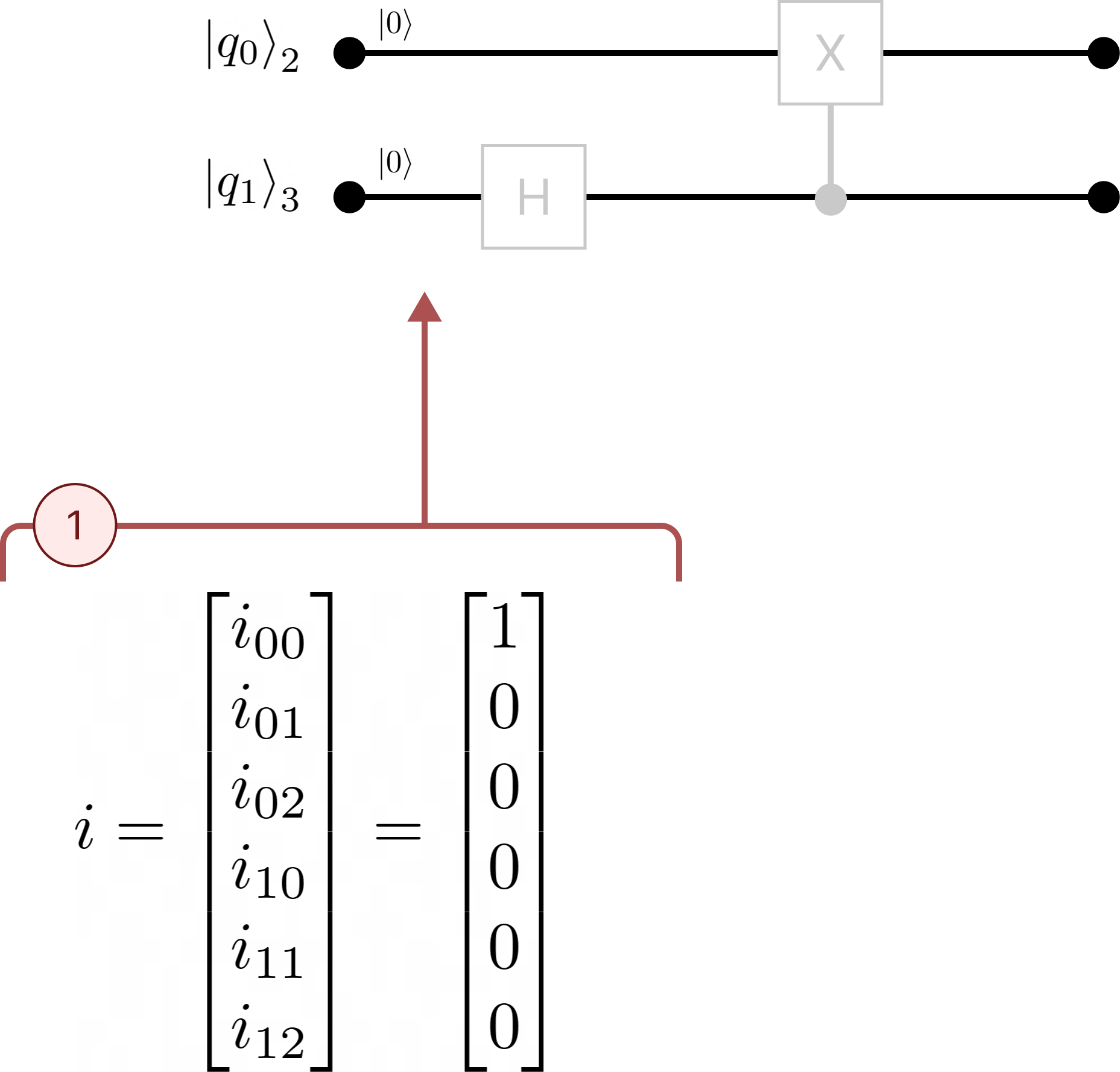}
    \caption{Step 1 of the example circuit execution: The statevector of the initial state is taken into account as a single block.}
    \label{fig:empty-state}
\end{figure}

Suppose, a circuit with 2 quantum registers $q_0$ and $q_1$, with dimensions $2$ and $3$, where we can apply a generalized Hadamard gate on depth $0$ at $q_1$, followed by a CX gate at depth $1$ with the target at $q_0$ and control at $q_1$. With an initial starting state of $|00\rangle$, the initial statevector, $|i\rangle$ will be as shown in Figure \ref{fig:empty-state}.

\begin{figure}[!h]
    \centering
    \includegraphics[width=.5\linewidth]{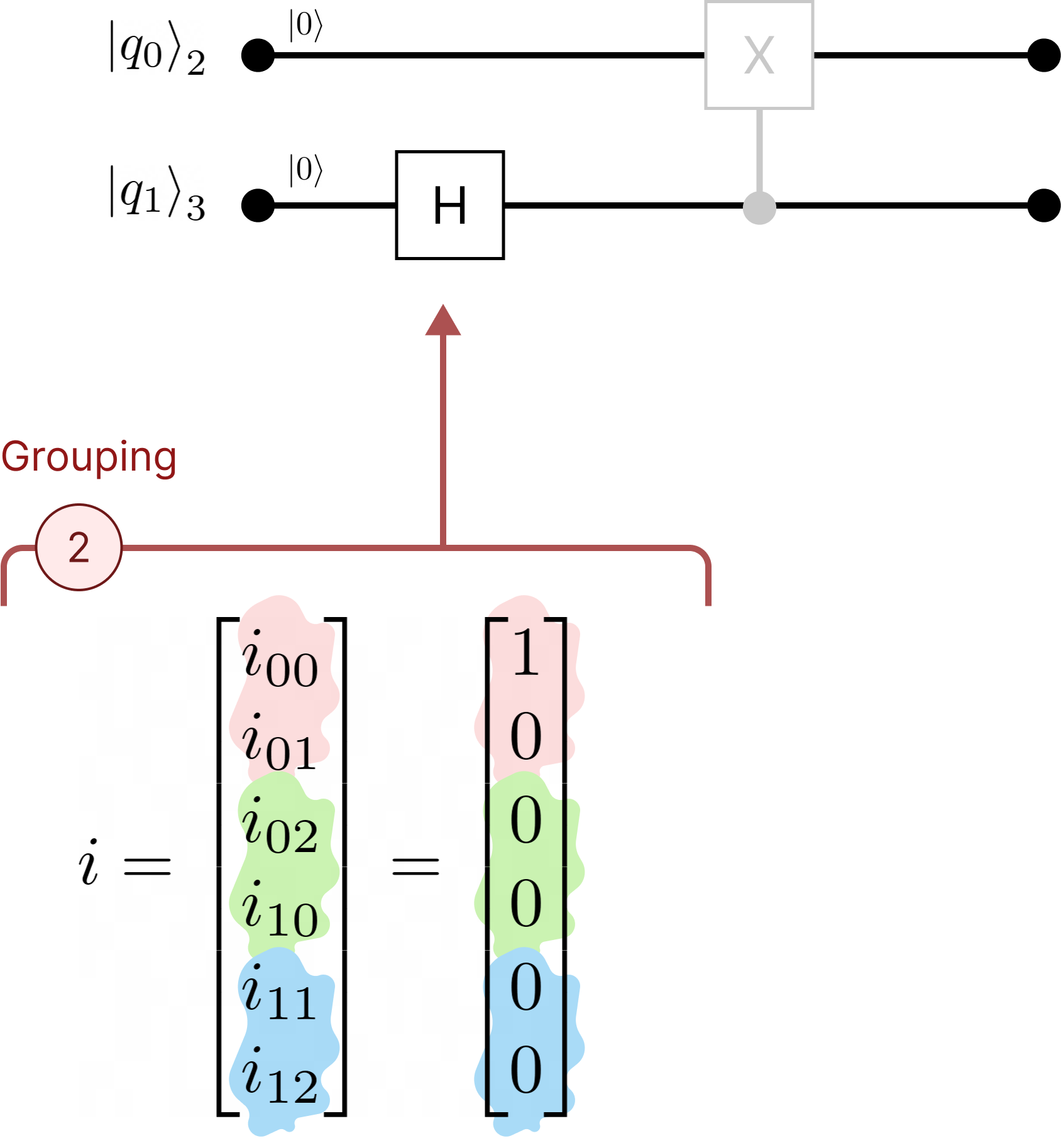}
    \caption{Step 2 of the example circuit execution: The statevector has been grouped into blocks based on the 3 dimensions of the qudit $q_1$.}
    \label{fig:hadamard-grouping}
\end{figure}

The simulator takes a 2-step approach (\textit{grouping} \& \textit{transformation}), for every single gate, to find the resultant statevector. The first generalized Hadamard gate applies on the 2nd qudit, i.e. $q_1$, so to transform the statevector as per the gate, the statevector components are needed to be grouped as per the qudit of interest, as shown in Figure \ref{fig:hadamard-grouping}. The statevector, in this case, has been divided into 3 groups, analogous to the 3 dimensions of the qudit of interest $q_1$. The groups are $\{i_{00}, i_{01}\}$, $\{i_{10}, i_{11}\}$, $\{i_{20}, i_{21}\}$. 

\begin{figure}[!h]
    \centering
    \includegraphics[width=.5\linewidth]{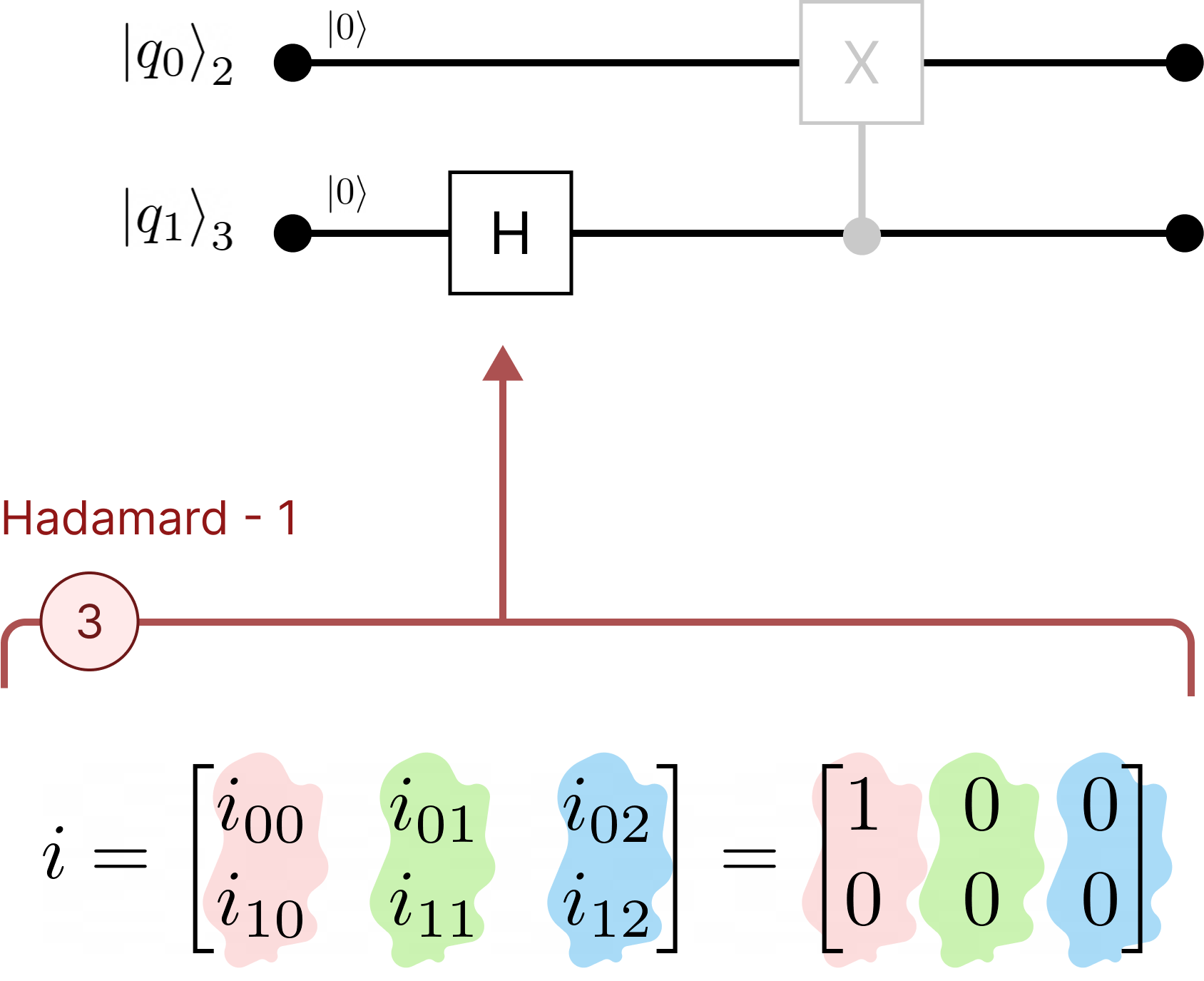}
    \caption{Step 3 of the example circuit execution: The grouped statevector is now being transformed into an appropriate matrix, such that further grouping is possible.}
    \label{fig:hadamard-transformation-alignment}
\end{figure}

This grouping empowers us to do logical manipulations and transformations, directly on the statevector. Once the groups are formed, the gate transformation can be applied to the groups. For the generalized Hadamard gate, the transformation would be superposition. This can be achieved by stacking the groups side-by-side as shown in Figure \ref{fig:hadamard-transformation-alignment}, for a better understanding of the operation. This side-by-side alignment allows the application of the Hadamard onto the corresponding statevector components.

\begin{figure}[!h]
    \centering
    \includegraphics[width=.5\linewidth]{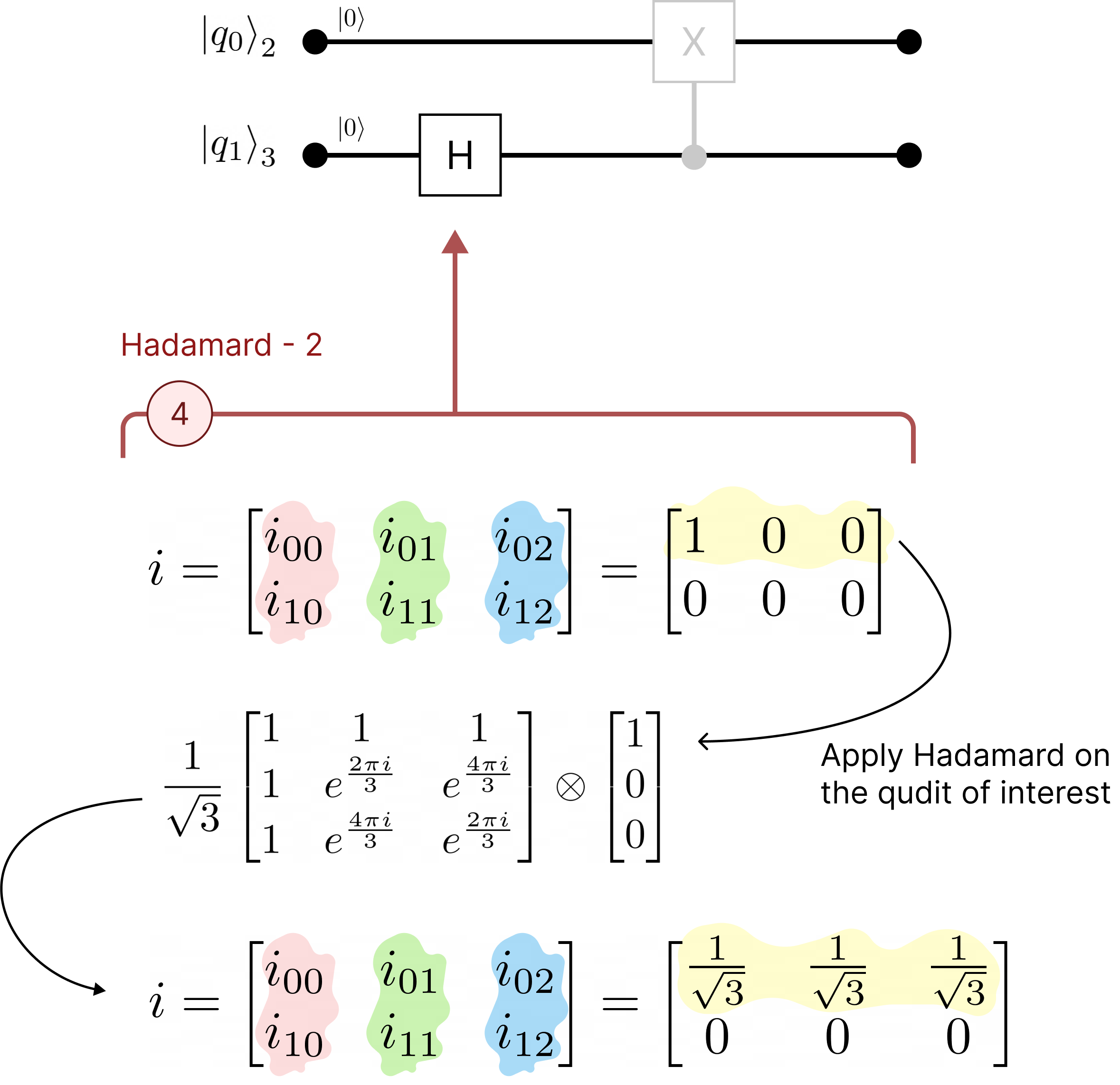}
    \caption{Step 4 of the example circuit execution: The Hadamard operator is now being applied to the statevector as illustrated above.}
    \label{fig:hadamard-transformation}
\end{figure}

 This alignment further helps to apply the generalized Hadamard transformation, in the correct places, as shown in Figure \ref{fig:hadamard-transformation}, which corresponds to the application of generalized Hadamard transformation, to be applied only on the required elements of the statevector, removing the necessity of maintaining exponentially big matrices and their costly operation.

\begin{figure}[!h]
    \centering
    \includegraphics[width=.5\linewidth]{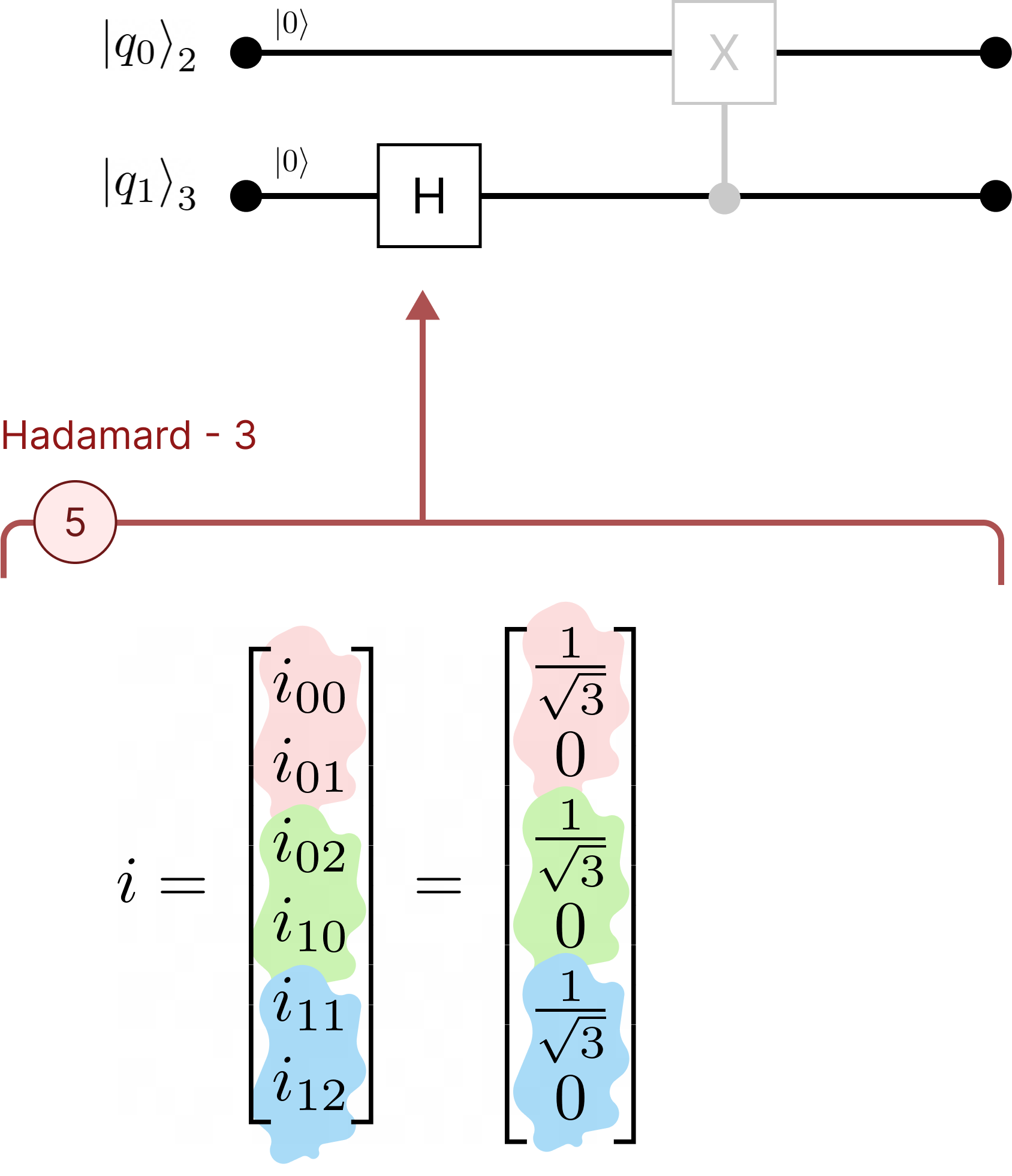}
    \caption{Step 5 of the example circuit execution: The resultant statevector at the end of the execution phase is now reconfigured into its initial form.}
    \label{fig:hadamard-transformation-realignment}
\end{figure}

Once the transformation for the generalized Hadamard is done, the statevector is realigned as shown in Figure \ref{fig:hadamard-transformation-realignment},  to get the columnar statevector back. This concludes the application of the generalized Hadamard gate, on the second qudit $q_1$. The above generalized Hadamard gate procedure can be summarized as a \textit{grouping} and \textit{shifting} procedure, over the existing transformation of the generalized Hadamard gate, thereby simplifying the overall computation. Following this kind of procedure, the gates can be applied to the statevector without concerning the rest of the qudits.

Following the generalized Hadamard gate, a CX gate is applied with the control placed in the second qudit $q_1$ and the target at the first qubit $q_0$.

For the CX gate, the same \textit{grouping} and \textit{shifting} procedure follows. First, the statevector components are grouped, based on the control of the CX gate, resulting in the very same grouping, as the generalized Hadamard gate. In the case of the CX gate, no shifting is required and can be applied directly.

Assuming that the CX gate increments the target qudit $q_0$ by $+1$ when the control qudit $q_1$ is at the highest possible state of $2$, the CX can be applied in the particular group of interest, based on this logic of CX gate. However, note that the control qudit $q_1$ needs to be at the highest possible state, implying that we will be dealing with the group $(i_{20}, i_{21})$ in the statevector, where $q_1 = 2$, i.e. the final group. The control qudit $q_0$ will be incremented by $+1$, implying that the elements in the group of interest, i.e. the final group, will shift its elements by $+1$.

\begin{figure}[!h]
    \centering
    \includegraphics[width=.5\linewidth]{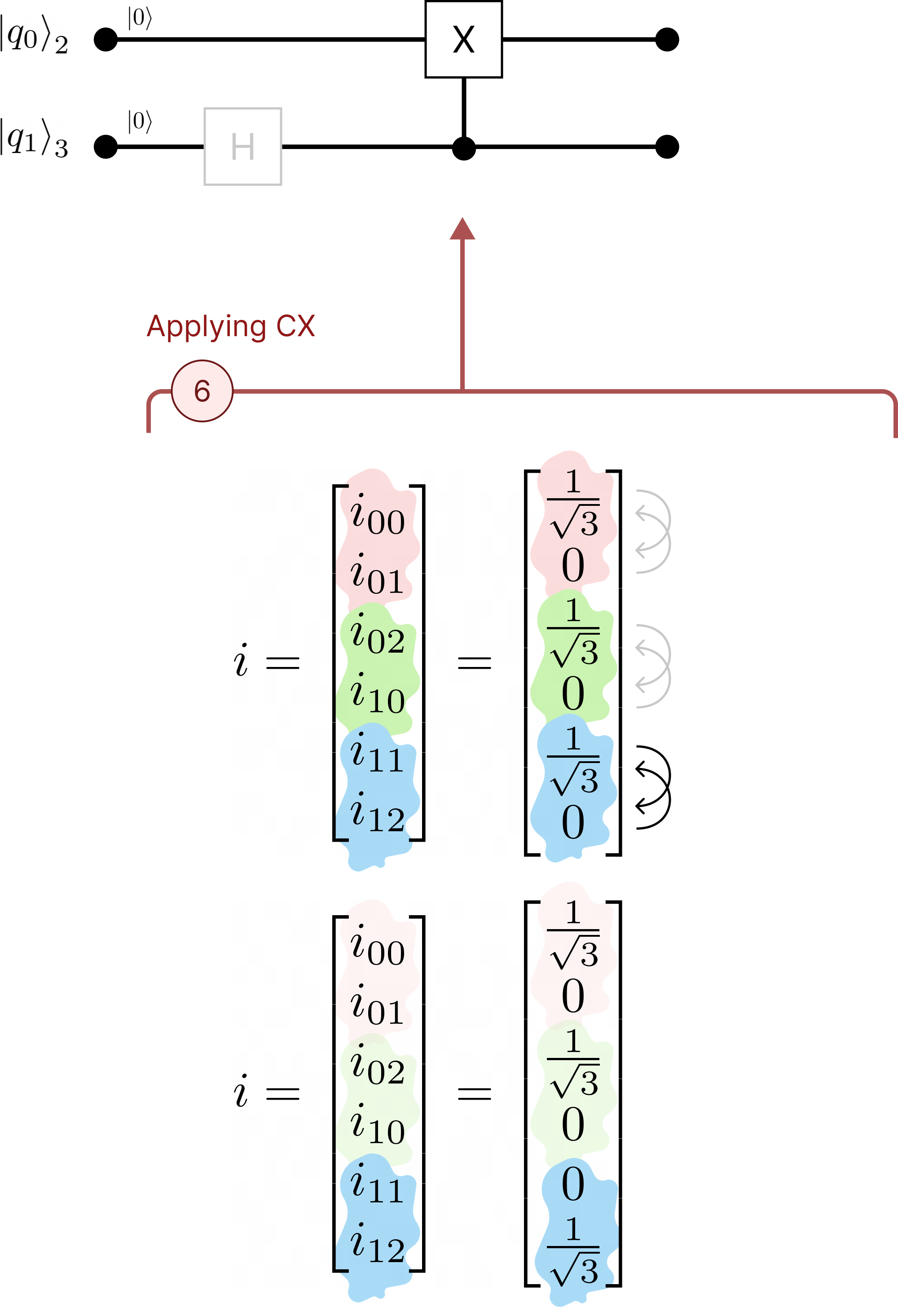}
    \caption{Step 6 of the example circuit execution: The CX gate is now applied to the resultant statevector as illustrated above.}
    \label{fig:cx-transformation}
\end{figure}

The main advantage of the proposed approach is that we have completely eliminated tensor products and matrix multiplications. The operations used in our approach are scalar multiplications, rotations, vector dot products, and vector additions, which are much more cheaper and efficient. Our approach does not require computing and storing the unitary matrix of size $D \times D$ where $D=\prod_{i=0}^{n-1}d_i$. We only store the statevector of size $D$, which is a quadratic improvement in the memory consumed. Further, there is also a future scope for parallelism across blocks as well as the repetitions of the blocks, which could provide significant advantages to multicore architectures. Moreover, the application of any unitary is achieved by inspecting the indices of non-zero amplitudes in the statevector, which is compatible with sparse implementations as well.

\section{Experimental Analysis}
For our method, we have utilized C++ for the implementation, in contrast to QuDiet which uses Python, as well as the Eigen library for faster linear algebra operations. 

The runtimes (in seconds) for some well-known quantum algorithms \cite{10.1145/3550488} using our method in contrast to QuDiet, Google Cirq, Microsoft Q\#, and IBM Qiskit are summarized in Table~\ref{tab:qubit_algos}. All experiments were run on a local computer with an 11th Gen Intel(R) Core(TM) i5-1135G7 processor operating at 2.40GHz, 2419 Mhz with 16 GB RAM, and a 64-bit Windows operating system.

\begin{table*}[!ht]
  \centering
  \caption{Comparative analysis of various circuits using our proposed methodology. Units for all numeric columns except the width and depth are in seconds.}
  \begin{tabular*}{\linewidth}{|p{0.02\linewidth}|p{0.08\linewidth}|p{0.04\linewidth}|p{0.04\linewidth}|p{0.05\linewidth}|p{0.073\linewidth}|p{0.071\linewidth}|p{0.07\linewidth}|p{0.055\linewidth}|p{0.05\linewidth}|p{0.07\linewidth}|p{0.074\linewidth}|}
  \hline 
      Sl. No. & Circuit & Width & Depth  & Cirq  & Microsoft Sparse & Microsoft full-state & IBM QASM & QuDiet & QuDiet Sparse & Our Method (Dense) & Our Method (Sparse)\\ \hline
      1 & Deutsch & 2 & 4  & 0.0039 & 0.0064 & 0.00652 & $2.025 \times 10^{-5}$ & 0.00075 & 0.0027 & $6.4701 \times 10^{-6}$ & $1.2504 \times 10^{-5}$\\ \hline
      2 & Grover & 2 & 11  & 0.0045 & 0.00605 & 0.00607 & $4.7385 \times 10^{-5}$ & 0.0023 & 0.0077 & $7.12 \times 10^{-6}$ & $5.7762 \times 10^{-5}$\\ \hline
      4 & HS4 & 4 & 11  & 0.0098 & 0.005983 & 0.00806 & $1.7658 \times 10^{-5}$ & 0.0042 & 0.0145 & $4.2987 \times 10^{-5}$ & $1.49515 \times 10^{-4}$\\ \hline
      5 & LPN & 5 & 4  & 0.0113 & 0.00578 & 0.0066 & $9.097 \times 10^{-5}$ & 0.0023 & 0.008 & $2.4716 \times 10^{-5}$ & $3.07076 \times 10^{-4}$\\ \hline
      6 & Simon & 6 & 13 & 0.0131 & 0.00497 & 0.0067 & $5.9843 \times 10^{-5}$ & 0.0055 & 0.0248 & $2.3641 \times 10^{-5}$ & $5.529 \times 10^{-5}$\\ \hline
      7 & Adder & 10 & 42  & 0.0189 & 0.0053 & 0.0105 & $6.9457 \times 10^{-5}$ & 2.035 & 2.91 & $4.27 \times 10^{-4}$ & $2.3393 \times 10^{-5}$\\ \hline
      8 & Multiplier & 10 & 55  & 0.0120 & 0.0054 & 0.0067 & $8.156 \times 10^{-5}$ & 1.78 & 0.288 & $2.42005 \times 10^{-4}$ & $1.5604 \times 10^{-5}$\\ \hline
      9 & SAT & 11 & 51  &  0.0139 & 0.0061 & 0.0087 & $5.9 \times 10^{-5}$ &196.156 & 34.87 & 0.00186 & $3.96242 \times 10^{-4}$ \\ \hline
      10 & GHZ state & 20 & 20   & 0.087 & 0.0058 & 0.224 & $5.283 \times 10^{-5}$& - & 0.1837 & 0.284126 & $1.4605 \times 10^{-5}$ \\ \hline
  \end{tabular*}
  \label{tab:qubit_algos} 
\end{table*}

\begin{figure}[!h]
    \centering
    \includegraphics[width=.5\linewidth]{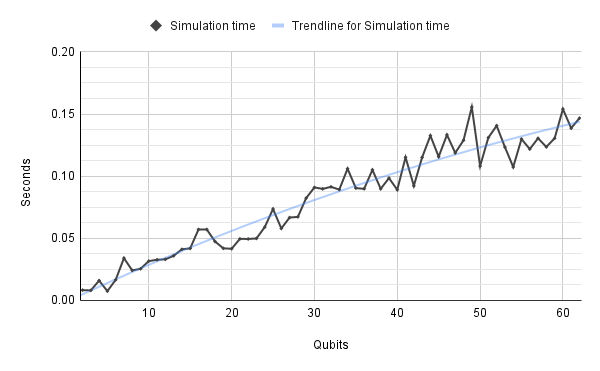}
    \caption{Simulation Time for GHZ for 1000 executions for (qubits)}
    \label{fig:sim_ghz_qubit}
\end{figure}

\begin{figure}[!h]
    \centering
    \includegraphics[width=.5\linewidth]{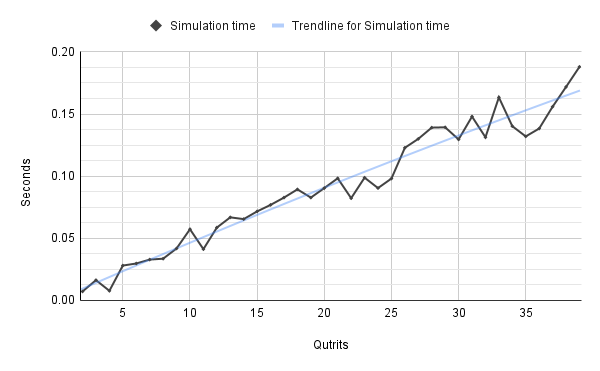}
    \caption{Simulation Time for GHZ for 1000 executions for (qutrits)}
    \label{fig:sim_ghz_qutrit}
\end{figure}

\begin{figure}[!h]
    \centering
    \includegraphics[width=.5\linewidth]{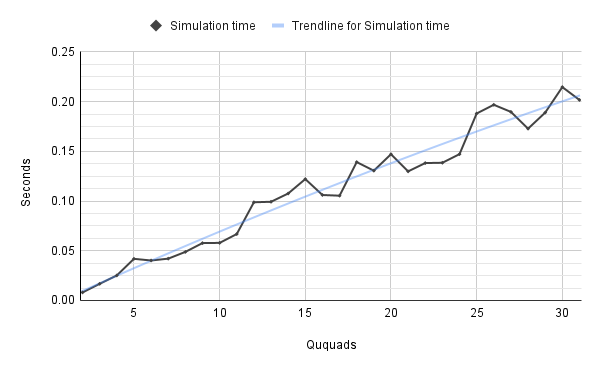}
    \caption{Simulation Time for GHZ for 1000 executions for (ququads)}
    \label{fig:sim_ghz_ququads}
\end{figure}

\begin{figure}[!h]
    \centering
    \includegraphics[width=.5\linewidth]{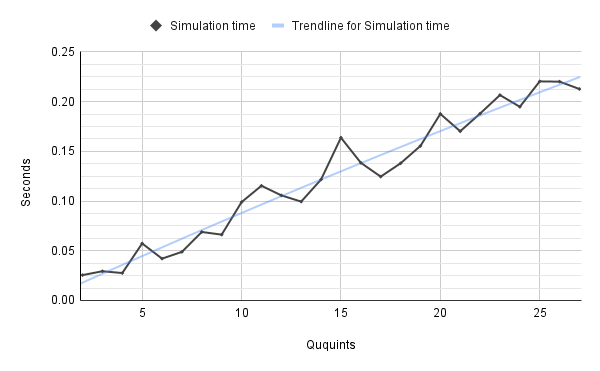}
    \caption{Simulation Time for GHZ for 1000 executions for (ququints)}
    \label{fig:sim_ghz_ququints}
\end{figure}

\begin{table*}[!ht]
    \centering
    
    \caption{Simulation time for qutrit circuits using our proposed methodology. Units for all numeric columns except the width and depth are in seconds.}
    \label{tab:qutrit}
    
    \begin{tabular}{|l|l|l|l|l|l|l|l|}
    \hline
        Sl. No. & Circuit & Width & Depth & QuDiet & QuDiet Sparse & Our Method (Dense) & Our Method (Sparse) \\ \hline
        1 & Deutsch & 2 & 4 & 0.0009 & 0.0042 & 7.27 $\times 10^{-6}$ & 2.45 $\times 10^{-5}$ \\ \hline
        2 & Grover & 2 & 11 & 0.003 & 0.0107 & 2.48 $\times 10^{-5}$ & 5.16 $\times 10^{-5} $\\ \hline
        4 & HS4 & 4 & 11 & 0.0128 & 0.0263 & 9.34 $\times 10^{-5}$ & 1.22 $\times 10^{-3}$ \\ \hline
        5 & LPN & 5 & 4 & 0.0165 & 0.016 & 6.13 $\times 10^{-5}$ & 1.27 $\times 10^{-3}$ \\ \hline
        6 & Simon & 6 & 13 & 0.434 & 0.039 & 1.47 $\times 10^{-4}$ & 2.74 $\times 10^{-4}$ \\ \hline
        7 & Adder & 10 & 42 & - & 1790.497 & 0.014 & 1.46 $\times 10^{-5}$ \\ \hline
        8 & Multiplier & 10 & 55 & - & 1.06 & 0.017 & 1.83 $\times 10^{-5}$ \\ \hline
        9 & SAT & 11 & 51 & - & - & 0.445 & 0.054 \\ \hline
        
    \end{tabular}
\end{table*}

\begin{table*}[!ht]
    \centering
    
    \caption{Simulation time for ququad circuits using our proposed methodology. Units for all numeric columns except the width and depth are in seconds.}
    \label{tab:ququad}
    \begin{tabular}{|l|l|l|l|l|l|l|l|}
    \hline
        Sl. No. & Circuit & Width & Depth & QuDiet & QuDiet Sparse & Our Method (Dense) & Our Method (Sparse) \\ \hline
        1 & Deutsch & 2 & 4 & 0.0013 & 0.0044 & 8.011  $ \times 10^{-6}$ & 4.1 x 10-5 \\ \hline
        2 & Grover & 2 & 11 & 0.0038 & 0.0116 & 2.62 $\times 10^{-5}$ & 7.7 $\times 10^{-5}$ \\ \hline
        4 & HS4 & 4 & 11 & 0.0412 & 0.099 & 1.73 $\times 10^{-4}$ & 3.38 $\times 10^{-3}$ \\ \hline
        5 & LPN & 5 & 4 & 0.498 & 0.2313 & 1.97 $\times 10^{-4}$ & 2.68 $\times 10^{-3}$ \\ \hline
        6 & Simon & 6 & 13 & 76.18 & 0.021 & 7.13 $\times 10^{-4}$ & 7.44 $\times 10^{-4}$\\ \hline
        7 & Adder & 10 & 42 & - & - & 0.55 & 1.49 $\times 10^{-5}$ \\ \hline
        8 & Multiplier & 10 & 55 & - & 22.452 & 0.424 & 2.08 $\times 10^{-5}$ \\ \hline
        9 & SAT & 11 & 51 & - & - & 5.597 & 0.066 \\ \hline
        
    \end{tabular}
\end{table*}

\begin{table*}[!ht]
    \centering
    
    \caption{Simulation time for ququint circuits using our proposed methodology. Units for all numeric columns except the width and depth are in seconds.}
    \label{tab:ququint}
    \begin{tabular}{|l|l|l|l|l|l|l|l|}
    \hline
        Sl. No. & Circuit & Width & Depth & QuDiet & QuDiet Sparse & Our Method (Dense) & Our Method (Sparse) \\ \hline
        1 & Deutsch & 2 & 4 & 0.0017 & 0.0045 & 8.964 $\times 10^{-6}$ & 6.32 $\times 10^{-5}$ \\ \hline
        2 & Grover & 2 & 11 & 0.0038 & 0.0114 & 3.32 $\times 10^{-5}$ & 8.85 $\times 10^{-5}$ \\ \hline
        4 & HS4 & 4 & 11 & 0.2523 & 0.954 & 3.13 $\times 10^{-4}$ & 1.16 $\times 10^{-2}$ \\ \hline
        5 & LPN & 5 & 4 & 10.23 & 4.393 & 5.53 $\times 10^{-4}$ & 6.8 $\times 10^{-3}$ \\ \hline
        6 & Simon & 6 & 13 & 4763.45 & 2.76 & 2.18 $\times 10^{-3}$ & 1.29 $\times 10^{-3}$ \\ \hline
        7 & Adder & 10 & 42 & - & - & 2.48 & 1.51 $\times 10^{-5}$ \\ \hline
        8 & Multiplier & 10 & 55 & - & 237.8075 & 2.45 & 1.85 $\times 10^{-5}$ \\ \hline
        9 & SAT & 11 & 51 & - & - & 51.564 & 0.34 \\ \hline
    \end{tabular}
\end{table*}

We also present the runtime trends for simulating GHZ state circuits on qubits, qutrits, ququads and ququints using the sparse matrices for our method in Fig. ~\ref{fig:sim_ghz_qubit}, ~\ref{fig:sim_ghz_qutrit}, ~\ref{fig:sim_ghz_ququads}, ~\ref{fig:sim_ghz_ququints} respectively. All of these graphs show a near-linear trend at which the runtime scales. This is because of the very few nonzero amplitudes that we deal with in GHZ state circuits and the linear increase in circuit depth. 

In the case of GHZ state circuit on $N$ qubits, at any point in time, we deal only with at most two nonzero amplitudes in the entire statevector of size $2^N$. Since the depth of the circuit is $N$, the runtime also scales linearly with $N$, due to the execution of $N$ gates sequentially. As we move to higher dimensional qudits, the number of nonzero amplitudes increases proportionally to the dimension of the qudit such as $3$ in the case of qutrits, $4$ for ququads, $5$ for ququints and so on.

We were able to efficiently leverage the sparse nature of these matrices to simulate GHZ state circuits on $62$ qubits in roughly $0.15$ milliseconds, $39$ qutrits in $0.188$ milliseconds, $31$ ququads in $0.2$ milliseconds and $27$ ququints in about $0.21$ milliseconds. The only reason for not simulating beyond these numbers was due to the size of the long long datatype used to store the basis states in C++ ($64$ bytes). 

\subsection{Comparative Analysis}
Table \ref{tab:qubit_algos} presents the runtimes of QuDiet and our proposal for various well-known quantum algorithms. It is evident that this method is much more scalable and memory-efficient than QuDiet. In fact, the method presented in this paper offers over $2\times$ improvement for the SAT circuit on $11$ qubits.

In tables, \ref{tab:qutrit} \ref{tab:ququad} \ref{tab:ququint}, we also present some higher and mixed-dimension quantum circuit simulation runtimes. From the data presented in these tables, we gather that the method presented therein is approximately 10$\times$ to 1000$\times$ faster than QuDiet for simulating multi-level quantum systems.

\section{Conclusion and Future Directions}
In this paper, we propose a higher-dimensional extension of the direct evolution approach to classical simulation of quantum circuits. The main advantage offered by this approach is a significant reduction in memory as well as the elimination of time-intensive and redundant operations such as tensor products and matrix multiplications. We also introduce a method, which implements the proposed block simulation approach in C++. A detailed analysis of the improvements offered by this method over its spiritual predecessor QuDiet, which is based on the traditional full-state simulation approach, is also presented for various qubit-qudit quantum systems as well as well-known quantum algorithms. Several large-scale simulations such as a $62$-qubit GHZ state circuit, etc. have been shown to be executed in just milliseconds using our proposed method. However, the scalability of this method and its sparse version is severely restricted by the datatype employed to store the basis states. We, therefore, plan to improve the scalability by employing datatypes with a larger range of values such as \textit{uint128} or even \textit{uint256} to run much larger circuits than presented in the benchmarks as the proposed simulation approach easily supports large-scale simulations. Several optimizations, such as the MCNOT gate implementation will be explored, which could further reduce the runtime. Hardware acceleration using GPUs and distributed execution \cite{farias2024quforgelibraryquditssimulation} are also important venues we wish to consider in the future. 

\section{Acknowledgment}
There is no conflict of interest.

\bibliographystyle{unsrt}  
\bibliography{biblio}

\begin{thebibliography}{10}

\bibitem{Chatterjee_2023}
Turbasu Chatterjee, Arnav Das, Subhayu~Kumar Bala, Amit Saha, Anupam Chattopadhyay, and Amlan Chakrabarti.
\newblock Qudiet: A classical simulation platform for qubit‐qudit hybrid quantum systems.
\newblock {\em IET Quantum Communication}, March 2023.

\bibitem{dalzell2023quantumalgorithmssurveyapplications}
Alexander~M. Dalzell, Sam McArdle, Mario Berta, Przemyslaw Bienias, Chi-Fang Chen, András Gilyén, Connor~T. Hann, Michael~J. Kastoryano, Emil~T. Khabiboulline, Aleksander Kubica, Grant Salton, Samson Wang, and Fernando G. S.~L. Brandão.
\newblock Quantum algorithms: A survey of applications and end-to-end complexities, 2023.

\bibitem{sahapra}
Amit Saha, Ritajit Majumdar, Debasri Saha, Amlan Chakrabarti, and Susmita Sur-Kolay.
\newblock Asymptotically improved circuit for a $d$-ary grover's algorithm with advanced decomposition of the $n$-qudit toffoli gate.
\newblock {\em Phys. Rev. A}, 105:062453, Jun 2022.

\bibitem{Wang_2020}
Y.~Wang, Z.~Hu, B.~C. Sanders, and S.~Kais.
\newblock Qudits and high-dimensional quantum computing.
\newblock {\em Frontiers in Physics}, 8, Nov 2020.

\bibitem{Ringbauer_2022}
Martin Ringbauer, Michael Meth, Lukas Postler, Roman Stricker, Rainer Blatt, Philipp Schindler, and Thomas Monz.
\newblock A universal qudit quantum processor with trapped ions.
\newblock {\em Nature Physics}, 18(9):1053–1057, July 2022.

\bibitem{qiskit2024}
Ali Javadi-Abhari, Matthew Treinish, Kevin Krsulich, Christopher~J. Wood, Jake Lishman, Julien Gacon, Simon Martiel, Paul~D. Nation, Lev~S. Bishop, Andrew~W. Cross, Blake~R. Johnson, and Jay~M. Gambetta.
\newblock Quantum computing with {Q}iskit, 2024.

\bibitem{cirq_developers_2024_11398048}
Cirq Developers.
\newblock Cirq, 2024.

\bibitem{Vincent_2022}
Trevor Vincent, Lee~J. Riordan, Mikhail Andrenkov, Jack Brown, Nathan Killoran, Haoyu Qi, and Ish Dhand.
\newblock Jet: Fast quantum circuit simulations with parallel task-based tensor-network contraction.
\newblock {\em Quantum}, 6:709, May 2022.

\bibitem{mato2024mqtquditssoftwareframework}
Kevin Mato, Martin Ringbauer, Lukas Burgholzer, and Robert Wille.
\newblock Mqt qudits: A software framework for mixed-dimensional quantum computing, 2024.

\bibitem{smelyanskiy2016qhipster}
Mikhail Smelyanskiy, Nicolas~PD Sawaya, and Al{\'a}n Aspuru-Guzik.
\newblock qhipster: The quantum high performance software testing environment.
\newblock {\em arXiv preprint arXiv:1601.07195}, 2016.

\bibitem{jones2019quest}
Tyson Jones, Anna Brown, Ian Bush, and Simon~C Benjamin.
\newblock Quest and high performance simulation of quantum computers.
\newblock {\em Scientific reports}, 9(1):10736, 2019.

\bibitem{jaques2022leveraging}
Samuel Jaques and Thomas H{\"a}ner.
\newblock Leveraging state sparsity for more efficient quantum simulations.
\newblock {\em ACM Transactions on Quantum Computing}, 3(3):1--17, 2022.

\bibitem{gauss1966disquisitiones}
Carl~Friedrich Gauss.
\newblock {\em Disquisitiones arithmeticae}, volume 157.
\newblock Yale University Press, 1966.

\bibitem{10.1145/3550488}
Ang Li, Samuel Stein, Sriram Krishnamoorthy, and James Ang.
\newblock Qasmbench: A low-level quantum benchmark suite for nisq evaluation and simulation.
\newblock {\em ACM Transactions on Quantum Computing}, 4(2), February 2023.

\bibitem{farias2024quforgelibraryquditssimulation}
Tiago de~Souza~Farias, Lucas Friedrich, and Jonas Maziero.
\newblock Quforge: A library for qudits simulation, 2024.

\end{thebibliography}

\end{document}